\pdfoutput=1
\documentclass[12pt]{article}
\usepackage{graphicx}
\usepackage{color}
\usepackage{bm}
\usepackage{amsfonts}
\usepackage{amssymb}
\usepackage{amstext}
\usepackage{amsmath}
\usepackage{fullpage}
\usepackage{url}
\usepackage[sort&compress,sectionbib]{natbib}
\def\cite#1{\citep{#1}}

\def\@{\partial}
\def\<{\langle}
\def\>{\rangle}
\def\nn{\nonumber}

\def\eg{{\it e.g.}}
\def\ie{{\it i.e.}}

\def\rb{\bar{r}}

\begin{document}
\title{
Nearly extensive sequential memory lifetime achieved by coupled nonlinear neurons
}
\author{
{Taro Toyoizumi}\\ 
{\it RIKEN Brain Science Institute}\\
{\it 2-1 Hirosawa, Wako, Saitama 351-0198, Japan}}
\date{\today}

\maketitle

\begin{abstract} 
  Many cognitive processes rely on the ability of the brain to hold
  sequences of events in short-term memory. Recent studies have
  revealed that such memory can be read out from the transient
  dynamics of a network of neurons. However, the memory performance of
  such a network in buffering past information has only been
  rigorously estimated in networks of linear neurons. When signal gain
  is kept low, so that neurons operate primarily in the linear part of
  their response nonlinearity, the memory lifetime is bounded by the
  square root of the network size. In this work, I demonstrate that it
  is possible to achieve a memory lifetime almost proportional to the
  network size, ``an extensive memory lifetime'', when the nonlinearity of
  neurons is appropriately utilized.  The analysis of
  neural activity revealed that nonlinear dynamics prevented the
  accumulation of noise by partially removing noise in each time
  step. With this error-correcting mechanism, I demonstrate that a
  memory lifetime of order $N/\log N$ can be achieved.
\end{abstract}

\section{Introduction}
Buffering a sequence of events in the activity of neurons is an
important property of the brain that is necessary to carry out many
cognitive tasks
\cite{Baddeley2000,Munte1998,deFockert2001,Orlov2000,Pastalkova2008,Hahnloser2002,Baeg2003}.
The fundamental limit of the capacity of the sequential memory is,
however, largely unknown. Several works have suggested that a long
memory lifetime can arise as a network property of neurons, where
individual neurons typically have limited memory
\cite{Goldman2009,White2004,Lim2011}. However, the structures and
operating regimes suitable for a network of neurons to buffer a
sequence of events is also unknown. This paper investigates the limit of
such sequential memory for buffering past stimuli in the presence of
dynamical noise. More specifically, we examine how reconstructions
of past stimuli degrade as we trace them back into the past. This kind
of working memory generally improves with the size of the
network. Hence, important questions are: How the memory lifetime
scales with network size, and what kind of network structure
achieves the longest memory lifetime.  The scaling of the memory
lifetime to the network size has been rigorously characterized only
under limited conditions \cite{White2004,Ganguli2008}. In particular,
the memory lifetime for non-saturating linear neurons can be
proportional to the network size, $N$, which is, from an information
theoretical perspective, the best possible situation for
reconstructing all sequences of non-sparse input
\cite{Ganguli2010}. This is called the extensive memory
lifetime. Ganguli et al. also estimated the memory lifetime of a
network of neurons with response nonlinearity but under a rather
restricted condition where the signal gain was kept small so that
neurons operated in their linear regime. Under this condition, the
sequential memory lifetime is upper-bounded by $\sim\sqrt{N}$
\cite{Ganguli2008}.  However, as we will see in the following,
fine-tuning
of a network parameter is necessary for this to work.

I explore in this paper a network structure that yields a long-lasting
sequential memory that is longer than the bound previously set for
nonlinear neurons.  The network structure that I explore is a simple
feedforward network with a fixed number of neurons in one layer. This
network architecture has been studied in the context of
synchronous-firing chains, \ie, synfire chains
\cite{Abeles1982,Bienenstock1995,Herrmann1995,Diesmann1999,Rossum2002,Vogels2005,Kumar2008}.
In the current context, reliable propagation of synfire activity is
used to maintain information on past sequences. Although reliable
propagation of synfire activity in the presence of noise has been
reported several times, quantitative characterization of such
reliability has been only partially achieved. In particular, previous
studies did not systematically evaluate the effect of occasional
strong noise that spontaneously ignites or blocks synfire activity
\cite{Herrmann1995,Bienenstock1995}. As we will see, this occasional
large noise prevents a network from achieving an extensive memory
lifetime.  The scaling of the memory lifetime to the network size in
the presence of such noise has not yet been reported to my knowledge.

In this paper, I analytically evaluate the effects of response
nonlinearity and noise on the performance of sequential memory. The
main result is the following: If we require a network of $N$
neurons to hold $I$ bits of information about stimulus presented at
each time, the achievable memory lifetime is proportional to
$(N/I)/\log(N/I)$, which is much longer than the previously proposed
order $\sqrt{N}$ memory, assuming a small gain.  Moreover, the
nonlinear dynamics of neurons drastically improves the
tolerance of working-memory to noise levels, compared to the
previously proposed semi-linear dynamical regime. Numerical
simulations show that complex firing sequences of leaky
integrate-and-fire neurons are successfully buffered by this network
architecture.

\section{Result}

In order to derive the memory lifetime of
feedforward networks, I consider a simple firing-rate model of
neurons with saturating response nonlinearity. We first aim at
reconstructing a sequence of binary input but we will show later that it
is straightforward to generalize this scheme to reconstruct sequences of 
analog input.

\subsection{Evaluating the memory lifetime for interacting nonlinear neurons}
\label{sec:NLchain}
\begin{figure}[h!]
\includegraphics[height=6cm,width=12cm]{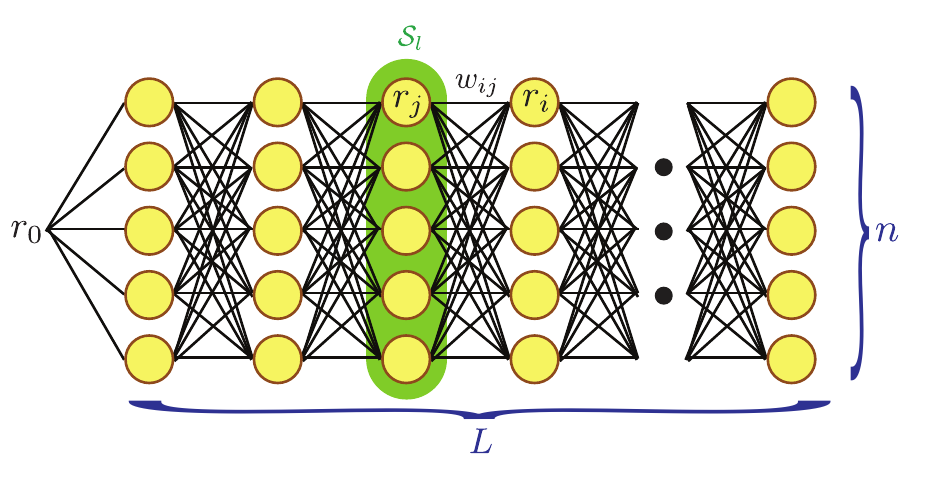}
\caption{\label{fig:ffnet} Simple feedforward network model of size
  $N$.  There are $n$ neurons in each layer, and there are $L=N/n$
  successive layers. Each neuron is connected to all the neurons in
  the previous layer with a uniform synaptic strength. We study how
  the input of strength, $r_0$, propagates down the feedforward
  chain.}
\end{figure}

To study what effects response nonlinearity has on the memory lifetime
in a simple system, we consider the dynamics of a homogeneous
feedforward network of $L$ layers (see, Fig.\ref{fig:ffnet}), where
each layer has $n$ neurons, and the total number of neurons in the
network is given by $N=n L$. Let us consider discrete-time dynamics
here for the sake of simplicity. The activity of neuron $i$ in layer
$l+1$ at time $t+1$ is modeled as
\begin{eqnarray}
  \label{eq:model}
  r_{i}(t+1)=\phi\left(\sum_{j\in \mathcal{S}_l}w_{ij}r_{j}(t)+\sigma \xi_{i}(t+1)\right),
\end{eqnarray}
where $\phi$ is the response nonlinearity, $\mathcal{S}_l$ is a set of
$n$ neurons in layer $l$, $w_{ij}=1/n$ is the uniform synaptic
strength from neuron $j$ in layer $l$ to neuron $i$ in layer $l+1$,
$\sigma$ is the magnitude of noise, and $\xi_{i}$ is an independent
white Gaussian random variable of unit variance that describes the
postsynaptic noise to neuron $i$. Here, in order to better distinguish
the effect of the nonlinearity from the signal-to-noise ratio, we fix the
magnitude of synaptic strength and instead change the slope of the
nonlinearity\begin{footnote}{We change the slope of $\phi$ by
    introducing scaling parameter $\beta$, with which the nonlinearity
    is written as $\phi(x)={\varphi}(\beta x)$ for some fixed
    nonlinearity $\varphi$.}\end{footnote} and the parameter
$\sigma$ that controls the noise level relative to the input from the
previous layer.  The time-dependent input to the network is described
by $r_0(t)$ and fed only to the first layer.  For simplicity, we
assume a sequence of binary input that takes either a positive or
negative value of the same magnitude.  Because the input to the
network at each time step propagates separately in this feedforward
model, we drop the time index, $t$, in the following and focus on the
propagation of a certain binary input signal, $r_0$. The information
about input degrades as the activity travels down the chain due to
noise. The task is to find the maximum number of
layers $L$ until which the information about the binary input reliably propagates.

\begin{figure}[h!]
\includegraphics[height=7cm,width=17cm]{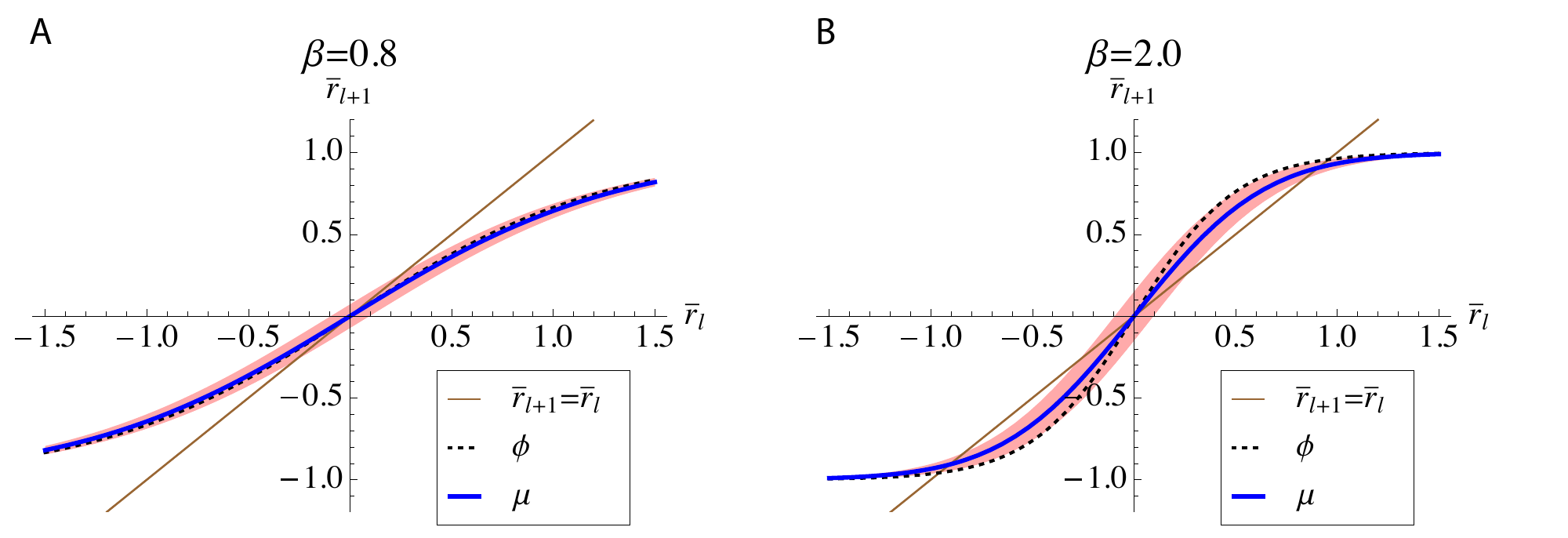}
\caption{\label{fig:rr} Conditional probability $P(\rb_{l+1}|\rb_l)$
  of the average activity.  The blue line is the conditional mean,
  $\mu(\rb_l)$, and the pink band is the conditional standard
  deviation, $\sqrt{v(\rb_l)/n}$.  The brown line indicates the
  condition $\rb_{l+1}=\rb_{l}$, and the black dashed line shows the
  nonlinear response function $\phi(x)=\tanh(\beta x)$.  (A) The slope
  of $\phi$ is $\beta=0.8$. Here, $\rb=\mu(\rb)$ has only one
  attracting solution at $\rb=0$. Hence, the activity tends to decay
  toward 0.  (B) The slope of $\phi$ is $\beta=2$. Here,
  $\rb_{l+1}=\mu(\rb_l)$ has three fixed points: two ($\rb\approx\pm
  0.9$) are attractive and one ($\rb=0$) is repulsive. When $\rb$ is
  close to one of the attracting fixed points, noise does not
  accumulate because it is partially removed at each time step. Other
  parameters are set to $\sigma=0.3$ and $n=10$.}
\end{figure}

Let $\rb_l\equiv\frac{1}{n}\sum_{i\in \mathcal{S}_l}r_{i}$ be the
average activity of the neurons in layer $l$. Because of the uniform
coupling strengths, $w_{ij}=1/n$, between adjacent layers the average 
activity of layer $l+1$ is given in terms
of the average activity of layer $l$ by
\begin{eqnarray}
  \label{eq:model2}
  \rb_{l+1}=\frac{1}{n}\sum_{i\in \mathcal{S}_{l+1}}\phi(\rb_{l}+\sigma \xi_{i}).
\end{eqnarray}
Note that Eq. \ref{eq:model2} is derived by averaging both sides of
Eq. \ref{eq:model} with $i\in \mathcal{S}_{l+1}$. Here, $\rb_{l+1}$ is the sum of
$n$ independent and identically distributed random variables.  Hence,
by using the central limit theorem, the conditional distribution, 
$P(\rb_{l+1}|\rb_l)$, approaches a Gaussian distribution
\begin{eqnarray}
  \label{eq:prob}
  P(\rb_{l+1}|\rb_{l})\approx \mathcal{N}\left(\mu(\rb_{l}),\frac{v(\rb_{l})}{n}\right)
\end{eqnarray}
for a large $n$. This distribution is characterized by the conditional mean,
$\mu(\rb_l)$, and the conditional variance, $v(\rb_l)/n$, calculated as:
\begin{eqnarray}
  \label{eq:mu_v}
  \mu(\rb)&=&\int \phi(\rb+\sigma \xi) D\xi,\nn\\
  v(\rb)&=&\int \phi^2(\rb+\sigma \xi) D\xi -\mu^2(\rb),
\end{eqnarray}
where $D\xi =\frac{e^{-\xi^2/2}}{\sqrt{2\pi}}d\xi$ describes a
Gaussian integral. These two quantities $\mu$ and $v$ are plotted in
Fig. \ref{fig:rr} for $\phi(x)=\tanh(\beta x)$.  With this
conditional probability, and for a given input $r_0$, the probability
distribution of the average activity in the final layer can be
formally described as $P(\bar{r}_L|r_0)=\int\cdots\int
P(\bar{r}_L|\bar{r}_{L-1})P(\bar{r}_{L-2}|\bar{r}_{L-3})\cdots
P(\bar{r}_1|r_0)d\bar{r}_{L}d\bar{r}_{L-2}\cdots d\bar{r}_1$, which is
sufficient to characterize the memory degradation at layer $L$.

In the following, let us consider a class of odd saturating
nonlinearity\footnote{More precisely, we consider a class of functions
  that satisfy: $\phi(x)=-\phi(-x)$ (odd), $|\phi(x)|\le 1$ (bounded),
  $\phi'(|x|)> 0$ (increasing), and $\phi''(|x|)< 0$ (saturating).},
such as $\phi(x)=\tanh(\beta x)$. For this class of functions, we can
show that the conditional mean, $\mu$, is also odd and
saturating. This means that the slope of $\mu$ is steepest at the
origin ($\mu'(x)\le \mu'(0)$ for $x\ne 0$).  Let us first consider a
trivial case, where the conditional variance, $v(\rb)/n$, is small and
negligible.  In this case, the dynamics is well approximated by a
deterministic update equation of the mean activity,
$\rb_{l+1}=\mu(\rb_l)$. Because $\mu(0)=0$ for odd $\phi$, $\rb=0$ is
always a fixed point in this dynamics. Let us call the slope $\mu'(0)$
 gain. If the gain is small ($\mu'(0)<1$), $\rb=0$ is a unique and
stable fixed point because $\mu$ is saturating. The average activity
must decay toward 0 (Fig. \ref{fig:rr}A). On the other hand, if the
gain is large ($\mu'(0)>1$), $\rb=0$ becomes an unstable fixed point,
and two stable fixed points (one positive and one negative) appear
(Fig. \ref{fig:rr}B). Hence, the average activity converges to either
the positive or the negative fixed point depending on the sign of the
input $r_0$. In general, when the conditional variance $v(\rb)/n$ is
not negligible, the activity fluctuates around the deterministic
dynamics described above, but the trend is similar. One important
property is that, while the conditional mean, $\mu(\rb)$, is
independent of the number of neurons per layer, $n$, the conditional variance
$v(\rb)/n$ decreases as $n$ increases. Hence, when
the gain is small ($\mu'(0)<1$), increasing the number of neurons in
each layer does not prevent the average activity from decaying. This
means that the memory lifetime is order $1$, \ie, the memory lifetime
does not scale with the number of neurons in each layer and is
determined by the gain. In contrast to the above case, when the gain
is large ($\mu'(0)>1$) and the activity approaches one of
the attracting fixed points, the memory degrades due to the
conditional variance, $v(\rb)/n$, and, hence, due to finite $n$. This
memory decay can be slowed down by increasing the number of neurons in
each layer. Even with additional noise at each time step, the
attracting force toward one of the stable fixed points can partially
remove this noise. This error-correcting dynamics that prevents noise
from accumulating becomes essential for the nearly extensive memory
lifetime as we will see in what follows.

Let us introduce an intuitive overview of how $\sim N/\log N$ memory
lifetime is derived. According to the attractor dynamics described
above, the average activity near an attracting fixed point can only be
driven closer to the other attracting fixed-point when uncommonly
large noise occurs. We estimate how often this rare flipping
occurs. The central limit theorem of Eq. \ref{eq:prob} states that the
effective noise level inversely decreases with the number of neurons
in each layer, $n$. Kramers' escape rate, \eg \cite{Risken1996},
yields that the probability of the activity flipping from one
attracting basin to the other in a particular layer is approximately
$e^{-n}$, where constant factors and higher order terms are
neglected. Hence, the probability that no flipping occurs throughout
$L=N/n$ successive layers is about $(1-e^{-n})^{L}\approx
\exp{\bm(}-(N/n)e^{-n}{\bm)}$, with which the input is correctly
estimated from the activity of the final layer. It is easy to see
that in order to keep this probability finite in the limit of large
$N$, $n$ should increase asymptotically faster than or equal to $\log
N$. Therefore, the best achievable memory lifetime is $L\sim N/\log N$.

Let us more rigorously evaluate a lower bound for the memory lifetime
when the dynamics is sufficiently nonlinear ($\mu'(0)>1$).  In this
case, we can choose a positive constant $r_c$ which satisfies
$\mu(r_c)> r_c$ (c.f. Fig. \ref{fig:rr}B). The basic strategy is to
assure with high probability that the average activity in the final
layer is $\rb_L\ge r_c$ if the input is also $r_0\ge r_c$.  Because of
the symmetry of the system, this condition also guarantees that the
activity is $\rb_L\le -r_c$ if the input is $r_0\le -r_c$.  Using the
Gaussian assumptions of $P(\rb_{l+1}|\rb_{l})$ for given $\rb_l$
(c.f. Eq. \ref{eq:prob}), the probability of $\rb_{l+1}\ge r_c$ is
expressed in terms of the error function by
\begin{eqnarray}
  \label{eq:error}
  P(\rb_{l+1}\ge r_c|\rb_{l}) &=& \int_{r_c}^{\infty} d\rb_{l+1} \, P(\rb_{l+1}|\rb_l)\nn\\
&=& 1-\frac{1}{2}\mbox{erfc}\left(\sqrt{\frac{n}{2}}\frac{\mu(\rb_{l})-r_c}{\sqrt{v(\rb_{l})}}\right).
\end{eqnarray}
Therefore, for any $\rb_{l}\ge r_c$, the probability of Eq. \ref{eq:error} is lower bounded by
 \begin{eqnarray}
   \label{eq:steperror}
   P(\rb_{l+1}\ge r_c|\rb_{l})\ge 1-\frac{1}{2}\mbox{erfc}\left(\sqrt{\frac{n}{2}}z_c\right),
 \end{eqnarray}
 where
 \begin{eqnarray}
   \label{eq:zc}
   z_c\equiv \min_{\rb_l\ge r_c}\frac{\mu(\rb_l)-r_c}{\sqrt{v(\rb_l)}}>\frac{\mu(r_c)-r_c}{\sqrt{1-\mu^2(r_c)}} >0
 \end{eqnarray}
 is a positive constant because $\mu(r_c)>r_c$.
 To obtain the first inequality in Eq. \ref{eq:zc} we used two
 properties: the variance of Eq.~\ref{eq:mu_v} is upper bounded by
 $v(\rb)\le 1-\mu^2(\rb)$ if $|\phi(x)|<1$, and $\mu$ is monotonically
 increasing with monotonically increasing $\phi$ (because
 $\mu'(r)=\int \phi'(r+\sigma\xi)D\xi>0$).  The right hand side of
 Eq. \ref{eq:steperror} can take a value close to 1 for large $n$
 ($>2/z_c^2$), suggesting that the average activity tends to remain in
 the same interval ($\rb\ge r_c$) as the previous layer with high
 probability.  If we assume that the input to the first layer is
 $r_0>r_c$, the probability that the average activity will reliably
 propagate through all layers without ever escaping below $r_c$ is
\begin{eqnarray}
  \label{eq:pc}
  P_c&\equiv& P(\{\rb_l\ge r_c\}_{l=1}^L|r_0)  \nn\\
&=& \int_{r_c}^\infty\cdots \int_{r_c}^\infty \prod_{l=1}^{L} \left[P(\rb_{l}|\rb_{l-1}) d\rb_{l} \right]\nn\\
  &\ge&\prod_{l=1}^L \left[1-\frac{1}{2}\mbox{erfc}\left(\sqrt{\frac{n}{2}}z_c\right)\right]\nn\\
&=& \left[1-\frac{1}{2}\mbox{erfc}\left(\sqrt{\frac{n}{2}}z_c\right)\right]^L
\end{eqnarray}
where Eq. \ref{eq:steperror} is used in the third line. To guarantee
a certain level of reliability, $P_c$, at the end of the chain, the
length of the chain, $L$, must be restricted by Eq. \ref{eq:pc}; \ie,
the length of the chain is at most
\begin{eqnarray}
  \label{eq:L}
  L&=&\frac{\log{P_c}}{\log\left[1-\frac{1}{2}\mbox{erfc}\left(\sqrt{\frac{n}{2}}z_c\right)\right]}\nn\\
&\approx& \frac{-2\log{P_c}}{\mbox{erfc}\left(\sqrt{\frac{n}{2}}z_c\right)}\nn\\
&\approx& C \sqrt{n} \; e^{\frac{n}{2}z_c^2},
\end{eqnarray}
where $C\equiv-\sqrt{2\pi}z_c \log{P_c}$ and, in the last two lines, higher
order terms are neglected assuming a large $n$.  Thus, the number of
layers where activity can reliably propagate increases as the number
of neurons in each layer increases. There is a constraint, on the
other hand, on the total number of neurons in the network, \ie,
\begin{eqnarray}
  \label{eq:N}
  N=n L= C n^{3/2}  \, e^{\frac{n}{2}z_c^2},
\end{eqnarray}
where the second equality follows assuming Eq. \ref{eq:L}.  For large
$N$, this equation yields to the leading order, $n\sim
(2/z_c^2)\log N$. Therefore, a memory lifetime of
\begin{eqnarray}
  \label{eq:L2}
  L=\frac{N}{n}\sim \frac{z_c^2}{2} \frac{N}{\log N}
\end{eqnarray} 
can be achieved with $n\sim (2/z_c^2)\log N$ neurons per layer.
Because of the symmetry of the system, we can repeat a similar
argument for $r_0<-r_c$ and find that the scaling is the same.  The
proportionality factor $z_c^2$ in Eq. \ref{eq:L2} describes the
signal-to-noise ratio. This factor takes a small value if $r_c$ is
small compared with the noise, showing that there is a minimum input
intensity for the network to store sequential memory reliably.
Although Eq. \ref{eq:L2} is a
lower bound for the memory lifetime as the inequality in
Eq. \ref{eq:pc} is not necessarily tight, we can expect that the
derived scaling behavior to $N$ is correct. This is because we can
also upper bound the probability of Eq. \ref{eq:error} using the same
functional form as Eq.~\ref{eq:steperror} but with another constant
factor greater than $z_c$.  The derived scaling of the memory lifetime
of order $N/\log N$ is much better than the previously
suggested \cite{Ganguli2008} scaling of order $\sqrt{N}$ for large $N$.

Although only a limited amount of information (at most 1 bit) can be
transmitted by the above network, it is easy to increase the amount of
information through the parallel use of $k$ chains.  Provided there is
independent input to each chain, the information transmitted through
the parallel chains becomes $k$ times larger than that through a
single chain.  While this solution requires $k$ times more neurons
than a single chain, this does not alter the scaling of the memory
lifetime to $N$. Therefore, the memory lifetime for reliably
reconstructing sequences of $\sim k$ bits of information in each time
step is $\sim(N/k)/\log (N/k)$.  More quantitatively, based on the
assumption that the input to each chain independently takes a positive
or negative value of the same magnitude with equal probability, the
mutual information between the input and the average activity in the
final layer is, by symmetry,
\begin{eqnarray}
  \label{eq:MI_NL}
  I(r_0;\rb_L) = k (1-H_2(P_c))
\end{eqnarray}
in bits, where the noise entropy, $H_2(P_c)\equiv -P_c\log_2 P_c
-(1-P_c)\log_2 (1-P_c)$, is about $0.5$ bits if $P_c=0.9$. This means that the
nonlinear feedforward chains can sustain $I$ bits of information about
the input for a duration proportional to $(N/I)\log (N/I)$.

\subsection{Numerical verification of the nearly extensive memory lifetime}

As an example, let us consider the sign nonlinearity,
$\phi(x)=\mbox{sgn}(x)$.  The corresponding conditional mean is given
by $\mu(\rb)=\mbox{erf}\left(\frac{\rb}{\sqrt{2}\sigma}\right)$. With
these binary neurons, where each neuron takes either an active
($r=+1$) or inactive ($r=-1$) state, it is easy to numerically
evaluate the memory lifetime because the average activity $\rb$ in
each layer can only take $n+1$ discrete values. For example, when $m$
$(m=0,1,\dots,n)$ neurons in layer $l+1$ are active and $n-m$ neurons
are inactive, the average activity in this layer is
\begin{eqnarray}
  \label{eq:rbm}
\rb_{l+1}=\frac{m}{n}-\frac{n-m}{n}=
\frac{2m-n}{n}.
\end{eqnarray}
Hence, the
conditional probability distribution is given in terms of a binomial distribution by
\begin{eqnarray}
  \label{eq:bin_condp}
P\left(\left.\rb_{l+1}=\frac{2m-n}{n}\right|\rb_l\right)=\binom{n}{m}
\left(\frac{1+\mu(\rb_l)}{2}\right)^{m}
\left(\frac{1-\mu(\rb_l)}{2}\right)^{n-m},
\end{eqnarray}
where $(1+\mu(\rb_l))/2$ and $(1-\mu(\rb_l))/2$ are the respective
probabilities that a neuron in layer $l+1$ will take an active or inactive
state. The conditional distribution of
Eq. \ref{eq:bin_condp} over all possible input/output states can be
described as an $n+1$ by $n+1$ square matrix. In particular, the
distribution of average activity in the final layer, $P(\rb_L|r_0)$,
can be computed by evaluating the $L$th power of this matrix. We
assume a decoder that estimates the sign of $r_0$ based on the sign of
$\rb_L$. This means that the performance is good if
$P(\rb_L>0|r_0)\approx 1$ for positive $r_0$; and this
condition also assures $P(\rb_L<0|r_0)\approx 1$ for negative $r_0$ by the symmetry.

\begin{figure}[h!]
\includegraphics[height=6cm,width=17cm]{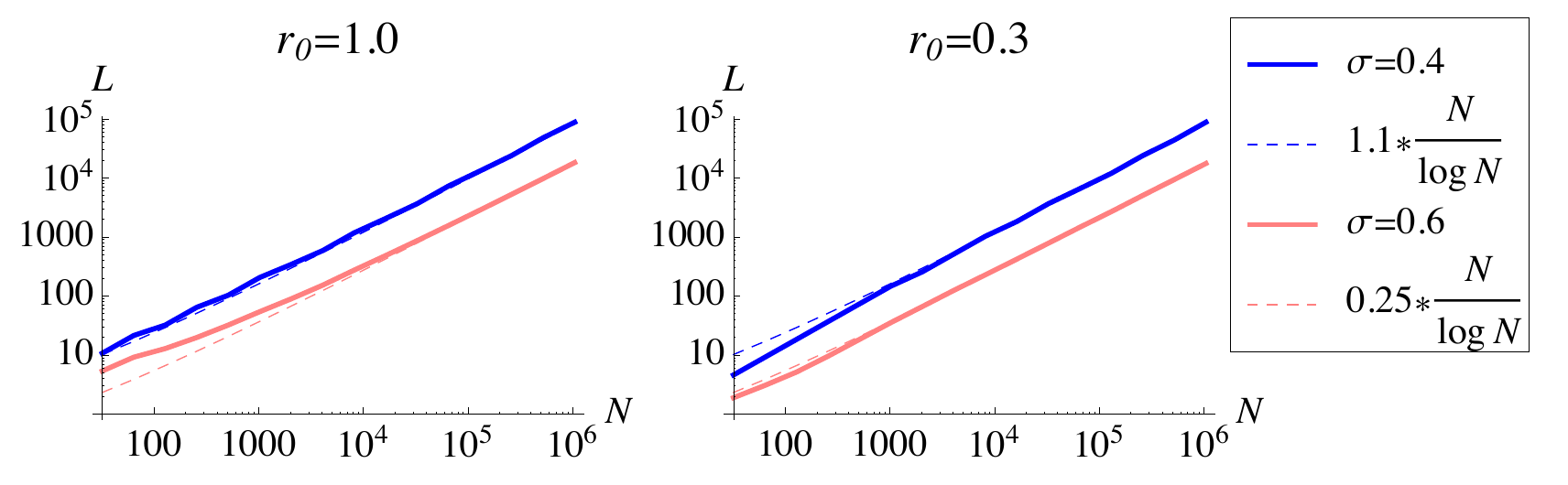}
\caption{\label{fig:simulation} Memory lifetime of binary neurons is
  scaled close to the network size. The memory lifetime, $L$, was
  evaluated at two noise levels: $\sigma=0.4$ (blue solid) and
  $\sigma=0.6$ (red solid), and two inputs: $r_0=1.0$ ({\it Left}) and
  $r_0=0.3$ ({\it Right}). The offset of two curves at different noise
  levels reflects the different number of neurons in each layer, $n$,
  chosen to achieve the 90\% decoding criterion.  The scaling behavior
  was well fitted by $\sim N/\log N$ in all cases as suggested by the
  theoretical result.  }
\end{figure}

Figure \ref{fig:simulation} plots the number of layers $L$, beyond
which probability $P(\rb_L>0|r_0)$ falls to less than 90\% at two
  different noise levels, $\sigma=0.4$ and $0.6$. The number of neurons in each
layer, $n$, was chosen to maximize the memory lifetime under a
constraint that the total number of neurons is $N$. We used two inputs,
$r_0=1.0$ and $r_0=0.3$, in the simulation, but the scaling of the
memory lifetime was not sensitive to the input $r_0$.
 We can see from Fig. \ref{fig:simulation} that the memory
lifetime is asymptotically proportional to $N/\log N$ as predicted by
the theory. Note that if the input is too small compared to the noise
level ($r_0\ll \sigma$), the asymptotic behavior is apparent only at
very large $N$ because a large number of neurons ($\gg
\sigma^2/r_0^2$) are required to achieve the 90\% reliability
criterion even in the first layer.

\subsection{Nonlinear dynamics provides robust working memory.}
We saw in the previous section that a nearly extensive memory
lifetime can be achieved by utilizing the error-correcting property of
nonlinear neural dynamics. In this section, I will show that the
sequential memory in this regime is much more robust to network
parameters than the previously proposed solution in the semi-linear
regime \cite{Ganguli2008}.

To better compare the sequential memory in a nonlinear vs. semi-linear
regime, let us clarify our goal. The goal is to maximize the length
of the feedforward chain(s), while maintaining $I$ bits of mutual
information about the input until the end of the chain(s). In this
section we consider that the input, $r_0$, is randomly drawn from a
Gaussian distribution of mean zero and variance $\sigma_0^2$. Because
the information about the input can only degrade as activity
propagates down the layers, it suffices to constrain the information
in the final layer, \ie, the mutual information between the input and
the average activity in the final layer must satisfy
\begin{eqnarray}
  \label{eq:MI1}
  I(r_0;\rb_L)\ge I.
\end{eqnarray}
Let us now review the sequential memory in the semi-linear regime
\cite{Ganguli2008}. The feedforward network structure in this
setting is similar to the one used in this paper except that the
number of neurons in each layer, $n_l$ for layer $l$, can vary.  The
total number of neurons is given by $N=\sum_{l=1}^L n_l$. The
derivation of Eqs. \ref{eq:prob} and \ref{eq:mu_v} is analogous with
the variable number of neurons in each layer. When the activity is
small, the conditional mean of Eq. \ref{eq:mu_v} is well approximated
by a linear function with slope (gain) $\mu'(0)$. As previously
explained, if the gain is smaller than 1, the signal tends to decay
toward zero, and if the gain is larger than 1, the signal tends to
grow until nonlinearity eventually kicks in
(c.f. Fig. \ref{fig:rr}). The optimal semi-linear solution is achieved
by setting the gain equal to 1 so that the signal neither decays nor
grows on average, and memory only degrades due to fluctuation of the
activity. In this case, the memory lifetime can scale with $\sqrt{N}$
\cite{Ganguli2008}. To implement this semi-linear solution,
however, some fine-tuning is required. For example, Ganguli et
al. simply used $\phi(x)=\tanh(x)$ without explicitly considering the
effect of noise that reduces the gain \cite {Herrmann1995}. Because
the slope of $\phi(x)=\tanh(x)$ is always less than 1 except at the
origin, the gain must be strictly less than 1 in the presence of noise
($\mu'(0)=\int D\xi \phi'(\sigma \xi)<1$). This means that the signal
must decay at each time step by the factor $\mu'(0)<1$, and because
the gain is independent of the number of neurons, the memory lifetime
is indeed order 1 rather than $\sqrt{N}$. To implement the memory
lifetime of $\sim\sqrt{N}$, one needs to fine-tune parameters such as
the slope $\beta$ of nonlinearity $\phi(x)=\tanh(\beta x)$. The
solution of $\beta$ that yields $\mu'(0)=1$ deviates from $\beta=1$ as
the noise level increases, and this difference becomes apparent at
large $N$. One should also note that the fine-tuning of the slope,
$\beta$, only provides a linear relation locally at around zero
activity ($\rb\approx 0$). If the activity is large in magnitude, the
signal tends to decay due to nonlinear effects.

Suppose we fine-tuned the gain to 1 and assumed the linear
input-output relation (which is not true if the signal and/or noise is
large). It is easy then to estimate the mutual information between
input and output because $P(\rb_L|r_0)$ is approximately Gaussian.
The signal-to-noise ratio in the final layer is described by
$\sigma_0^2/(\sum_{l=1}^L \sigma^2/n_l)$, where the signal is
preserved by setting the gain equal to 1, and the noise of variance
$\sigma^2/n_l$ is added in each layer. Hence, under this semi-linear
scheme, the mutual information is
\begin{eqnarray}
  \label{eq:MI_SL}
  I(r_0;\rb_L)&=&\frac{1}{2}\log_2 \left[1+\frac{\sigma_0^2}{L\sigma^2}\left(\frac{1}{L}\sum_{l=1}^L \frac{1}{n_l}\right)^{-1}\right]\nn\\
&\le &\frac{1}{2}\log_2 \left[1+\frac{\sigma_0^2}{L\sigma^2}\left( \frac{1}{\frac{1}{L}\sum_{l=1}^L n_l}\right)^{-1}\right]\nn\\
&=&\frac{1}{2}\log_2 \left[1+\frac{N\sigma_0^2}{L^2\sigma^2}\right],
\end{eqnarray}
where the second line follows due to the convexity of the $1/x$ function, 
and the equality holds if and only if the number of neurons in each
layer is uniform ($n_l=N/L$ for all $l$) \cite{Lim2011}. Hence, the best achievable memory
lifetime that guarantees $I$ bits of information about input under
this semi-linear scheme is
\begin{eqnarray}
  \label{eq:L_SL1}
  L=\min\left(\frac{\sigma_0}{\sigma}\sqrt{\frac{N}{2^{2I}-1}},N\right),
\end{eqnarray}
where $\min$ takes the minimum argument. This means that unless the
number of neurons is small, \ie, $N<
(\sigma_0^2/\sigma^2)/(2^{2I}-1)$, the memory lifetime of the
semi-linear network is proportional to $\sqrt{N}$ and exponentially
decreases with $I$, suggesting that it is difficult to maintain
precise information about input in this setting. If a large amount of
information is required, however, we can apply the parallel scheme
used in Sec. \ref{sec:NLchain} to the semi-linear memory by dividing
$N$ neurons to $k$ parallel chains, where each chain consists of $N/k$
neurons. Provided there is independent input to each chain, a single chain only
needs to retain $I/k$ bits of information in the final layers. Hence, the
memory lifetime of the semi-linear parallel chains becomes
\begin{eqnarray}
  \label{eq:L_SL2}
  L=\min\left(\frac{\sigma_0}{\sigma}\sqrt{\frac{(N/k)}{2^{2I/k}-1}},\frac{N}{k}\right).
\end{eqnarray}
In Eq. \ref{eq:L_SL2}, the first and second arguments of $\min(\cdot)$ are 
increasing and decreasing functions of $k$, respectively. Hence, the
best memory lifetime for large $N$ is given by
\begin{eqnarray}
  \label{eq:L_SL3}
  L \approx \frac{\sigma_0}{\sigma}\sqrt{\frac{N}{(2\log 2)I}}
\end{eqnarray}
at $k\approx (\sigma/\sigma_0)\sqrt{(2\log 2) N I}$, where each chain
needs to retain much less information than the single chain scheme.
Although the use of parallel chains does not alter the asymptotic
scaling of memory lifetime to $N$, it is also beneficial in the
semi-linear regime to buffer a large amount of information. Comparing
the memory lifetime in two regimes, $\sim(N/I)\log(N/I)$ in the
nonlinear regime and $\sim\sqrt{N/I}$ in the semi-linear regime, we
can conclude that the memory lifetime to retain the same amount of
information increases asymptotically faster with $N$ in the nonlinear
regime than in the semi-linear
regime.

\begin{figure}[h!]
\includegraphics[width=16.5cm]{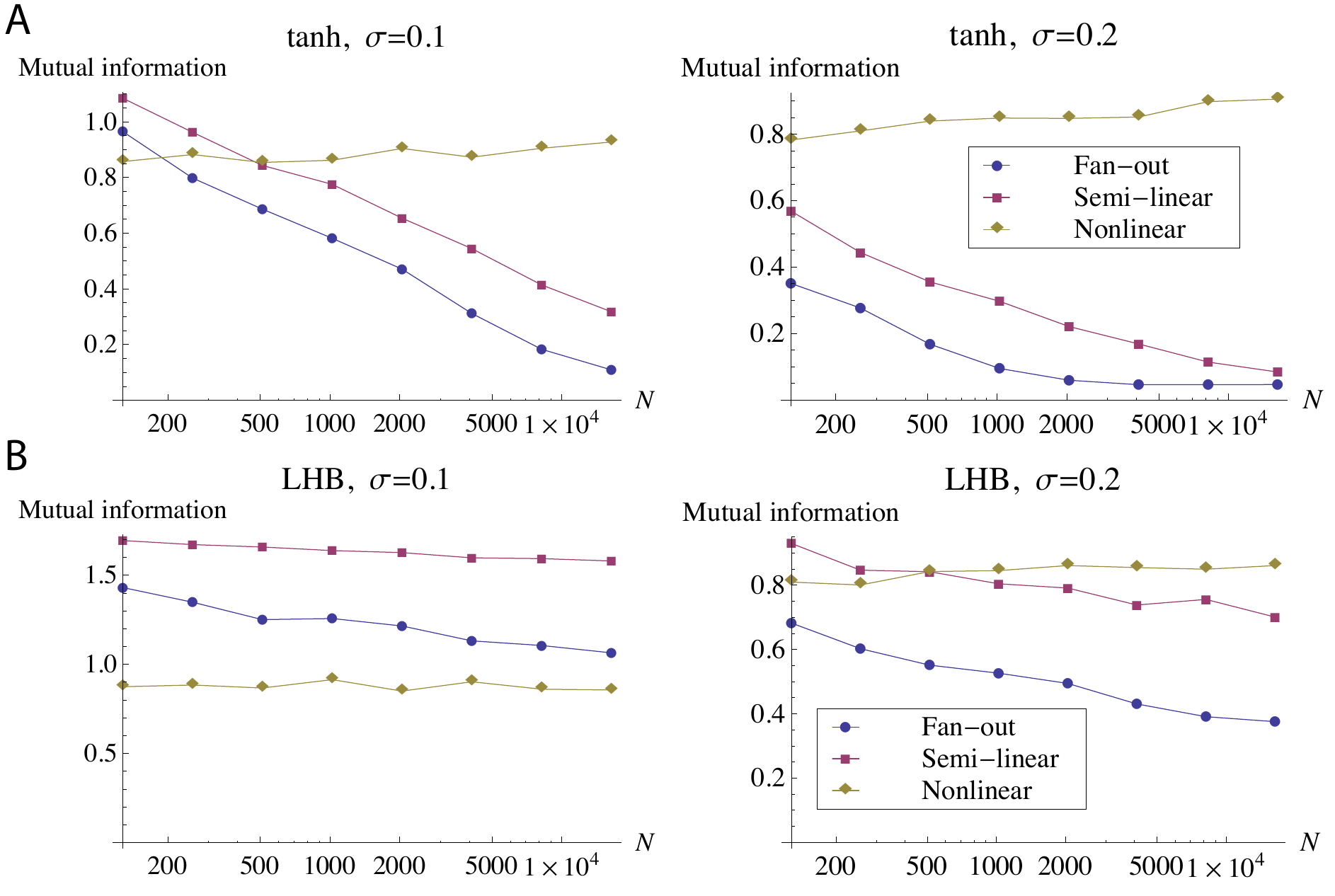}
\caption{\label{fig:GHS} The nonlinear network outperformed the
    semi-linear network for large network sizes or with large noise.
    A Gaussian random input with zero mean and standard deviation
    $\sigma_0=0.5$ was provided to the first layer, and the
    performance was measured by the mutual information of the input
    and the activity in the final layer. The sequential memory
    performance was numerically examined at noise level $\sigma$,
    shown in each panel and with two nonlinear functions: (A) the
    hyperbolic tangent nonlinearity $\phi(x)=\tanh(\beta x)$ and (B) a
    piecewise-linear function $\phi(x)=\beta x$ for $|x|<1/\beta$ with
    hard saturating bounds (LHB). Three types of networks were
    compared with approximately the same size, $N$, and the same
    number of layers $L\approx \sqrt{2N}$ for a fair comparison: the
    fan-out network \cite{Ganguli2008} with $\beta=1$ and a linearly
    increasing number of neurons along deeper layers ($n_l=l$; order 1
    memory lifetime); the semi-linear network with a
    $\beta$ solution that yielded a gain equal to 1 and a fixed number
    of neurons in each layer ($n_l=N/L$; order $\sqrt{N}$ memory
    lifetime); and the nonlinear network with the same
    network architecture as the semi-linear network but with $\beta=2$
    ($n_l=N/L$, order $N/\log N$ memory lifetime). The
    semi-linear network always showed better performance than the
    fan-out network and the nonlinear network was superior to the
    other two except at a small network size and with a small amount
    of noise.}
\end{figure}

In practice, the network size $N$ is always finite. Hence, whether we
see the differences in the order 1, $\sqrt{N}$, and $N/\log N$ memory
lifetimes depends on the network size. I therefore investigated three
different networks with about the same number of total neurons $N$ and
the same number of layers $L$: the fan-out network \cite{Ganguli2008}
with a linearly increasing number of neurons along layers ($n_l=l$)
and $\beta=1$; the semi-linear network with a fixed number of neurons
in each layer and a $\beta$ solution that yielded a gain equal to 1;
and the nonlinear network, which had the same network architecture as
the semi-linear network, but with $\beta=2$ (and a gain greater than
1). For a fair comparison, the number of neurons per layer was
adjusted for the semi-linear and nonlinear networks so that the total
number of neurons was approximately the same as that of the fan-out
network, \ie, $N=L(L+1)/2$.  Figure \ref{fig:GHS}A shows the mutual
information between the Gaussian input and the activity in the final
layer for the three networks introduced above for various network
sizes with $\phi(x)=\tanh(\beta x)$. When noise was small
($\sigma=0.1$), the performance of the semi-linear and fan-out
networks was superior to that of the non-linear network with a small
network size. This was because the nonlinear network squashed the
analog inputs to almost binary values, reducing information down to
about one bit, but the fan-out and semi-linear networks were able to
retain more than one bit of information at a small network size and
with low noise. As expected, the semi-linear network preserved more
information than the fan-out network because of the fine-tuning of
$\beta$ and the optimal network architecture.  When the network size
was larger than 500, the nonlinear network preserved more information
than the other networks.  At a slightly higher noise level
($\sigma=0.2$), the nonlinear network always outperformed the other
two in the range of network sizes examined. Note that the mutual
information of the nonlinear network increased with the network size
here because the number of layers was matched to the fan-out
network. This means that the nonlinear chain can be significantly
longer than the fan-out and semi-linear chains to achieve a comparable
level of mutual information. Figure \ref{fig:GHS}B shows results
analogous to Fig. \ref{fig:GHS}A but using for $\phi$ a linear
function with hard saturating bounds (LHB), \ie,
$\phi(x)=\mbox{sgn}(x)\min(|\beta x|,1)$. The results were
qualitatively similar to Fig. \ref{fig:GHS}A but the fan-out and
semi-linear networks performed better in this figure because LHB
retained linearity for a larger range of input than the hyperbolic
tangent function. In particular, compared at the same noise levels,
the crossover point of the semi-linear and nonlinear networks with LHB
nonlinearity lay at a larger $N$ than with the tangent hyperbolic
nonlinearity\footnote{Although the semi-linear network showed better
  performance than the nonlinear network for the entire range of $N$
  examined in Fig. \ref{fig:GHS}B with $\sigma=0.1$, the difference in
  the scaling of the memory lifetime ensures a crossover at a
  larger network size.}.  Because the nonlinear functions used in
\ref{fig:GHS}A and \ref{fig:GHS}B share the same slope at the origin,
the result shows that not only $N$ and $\sigma$, but also the
nonlinearity affects the crossover point of the
semi-linear and nonlinear networks.

We should also note that when more biological neuron models
are used, it is even more difficult for the semi-linear model to set
the gain equal to 1 because the nonlinearity is not fixed but changes
with the dynamical input properties in those models. Hence, the
semi-linear memory requires some elaborate additional mechanism to
achieve full performance.  Another important and potentially testable
difference between the semi-linear and non-linear memory is how
the memory lifetime, $L$, scales with the variance, $v$, of the
network activity, \ie, $L\sim v^{-\gamma}$ with some exponent
$\gamma$. While the semi-linear memory provides $\gamma= 1/2$ from
Eq. \ref{eq:L_SL2}, the non-linear memory provides $\gamma=1$ from
Eq.~\ref{eq:L2}.

\subsection{Synfire chains can reliably buffer complex spike sequences of leaky integrate-and-fire neurons.}

The abstract firing rate model studied in the previous sections was
suitable for mathematical analyses but was less biologically
realistic. However, all the main properties explored in the previous
sections should hold even with more realistic models. The key
properties that yielded the nearly extensive sequential memory
lifetime were the feedforward propagation of activity (that prevents
stimuli presented at different timings from mixing) and the attracting
dynamics (that implements error-correction). To
illustrate this point, a feedforward network of leaky integrate-and-fire (LIF)
neurons was explored. Detailed parameter studies were, however, not
the scope of this paper.

\begin{figure}[h!]
\includegraphics[height=7cm,width=17cm]{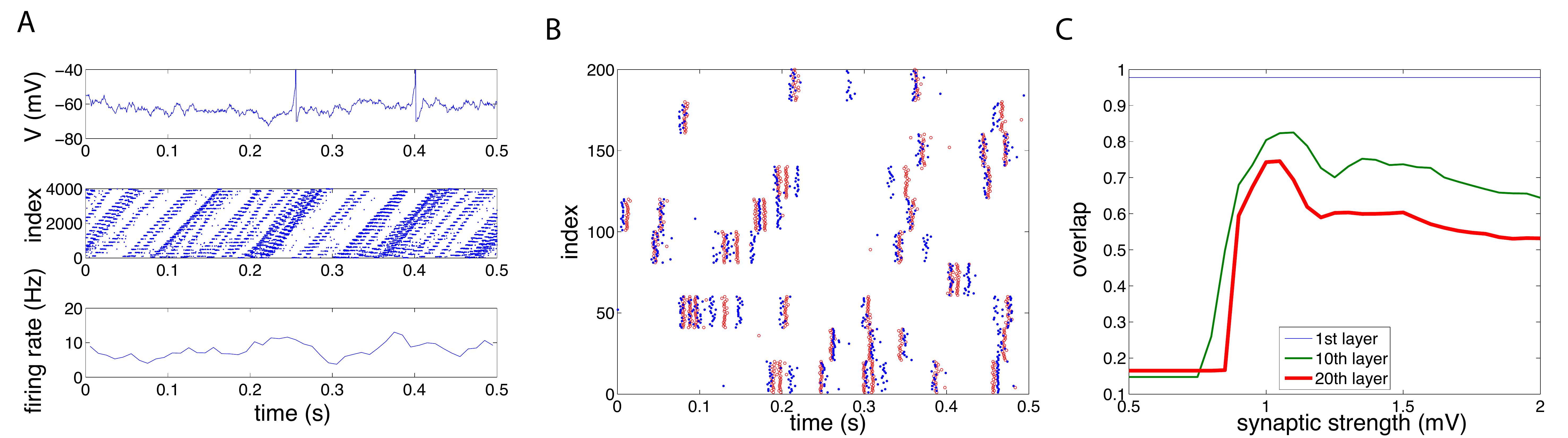}
\caption{\label{fig:LIF} A feedforward network of leaky
  integrate-and-fire neurons reliably buffered spike patterns. (A)
  {\it Top}: Membrane potential of a single neuron. {\it Middle}:
  Spike-timing of all neurons in the network. The network consisted of
  $k=10$ independent synfire chains, where each synfire chain had
  $L=20$ layers and $n=20$ neurons in each layer. The oblique patterns
  describe feedforward propagation of synfire activity. {\it Bottom}:
  The population firing rate of all the neurons averaged in 10-ms
  bins. (B) The spiking pattern of the first layers (blue dots) was
  well preserved even until the final 20th layers (red circles). The
  spike pattern of the final layers was shifted so that the spike
  overlap with the first layers was maximized. The feedforward
  synaptic strengths were set to 1 mV. Note that input pulses were
  somewhat degraded by background noise even in the first layer. (C)
  The spike overlap with the input pulses in the 1st, 10th, and 20th layers plotted for different feedforward
  synaptic strengths.}
\end{figure}

 A network of $N=4000$ current-based LIF
neurons was simulated, \eg, \cite{Dayan2001,Vogels2005}.  The
network consists of $k=10$ independent synfire chains, where each chain
has $L=20$ layers and $n=20$ neurons in each layer.  The membrane
dynamics of neurons $i \, (i=1,2,\dots,N)$ is described by
\begin{eqnarray}
  \label{eq:LIF}
  \tau\frac{dV_i}{dt}(t) = -V_i(t)+E_L + x_i(t) + \mu_{bg}+\sigma_{bg}\xi_i(t),
\end{eqnarray}
where $V_i$ is the membrane potential of neuron $i$, $\tau=10$ ms is
the membrane time-constant, $E_L=-70$ mV is the resting potential,
$x_i$ is the input to neuron $i$ from other neurons in a local
network, $\mu_{bg}=7$ mV is the mean background input level, $\xi_i$
is white Gaussian noise of unit variance, and $\sigma_{bg}=5$ mV
describes the magnitude of background input fluctuation. When the
membrane potential reaches the threshold value of $V_{th}=-50$ mV, the
neuron emits a spike and the membrane potential is reset to
$V_{reset}=-70$ mV. After the spike, the membrane potential is fixed
at $V_{rest}$ for the duration of the refractory period,
$\tau_{ref}=2$ ms.  Input to neuron $i$ is calculated according to
\begin{eqnarray}
  \label{eq:x}
  \tau_x \frac{dx_i}{dt}(t) = -x_i(t)+\sum_{j=1}^N \sum_{f} w_{ij}\delta(t-t_j^{(f)})
\end{eqnarray}
where $\tau_x=5$ ms, $w_{ij}$ is the synaptic strength from neuron $j$
to $i$, $\delta(t)$ is the Dirac delta function, and $t_j^{(f)}$ is
the $f$th spike time of neuron $j$.  This means that when neuron $i$
receives a spike from neuron $j$ at $t_j^{(f)}$, its membrane
potential is depolarized by
$w_{ij}[e^{-(t-t_j^{(f)})/\tau}-e^{-(t-t_j^{(f)})/\tau_x}]/(\tau-\tau_x)$
for $t>t_j^{(t)}$. We measure the synaptic strength $w_{ij}$ using the
peak amplitude of the excitatory postsynaptic potential, which is
about $w_{ij} * 50 \, \mbox{s}^{-1}$ for the current set of
parameters.  The synaptic strength, $w_{ij}$, takes a uniform non-zero
value, $w$, if the layer of neuron $i$ is next to the layer of neuron
$j$, and takes zero otherwise.  The feedforward synaptic strength
between adjacent layers is set uniformly to $1.0$ mV except in
Fig. \ref{fig:LIF}C, where $w$ is varied as a control parameter.
Each chain is independently stimulated
by $10$-Hz random Poisson pulses, upon which all the neurons in
the first layer are depolarized by $10$ mV according to
the time course of excitatory synaptic input.

Figure \ref{fig:LIF}A shows the model behavior. The top panel shows
the membrane potential of a single neuron. The middle panel shows the
spiking pattern of all neurons, where neurons were indexed first
within the same layer of the same chain, then across chains, and
finally across layers. The oblique arrangements of spiking patterns in
the middle panel demonstrates that most of the synchronous firing
patters evoked by external input pulses successfully reached the final
layers in about 100 ms. The bottom panel shows that the overall
population firing rate of the network was kept at about 10 Hz.  Figure
\ref{fig:LIF}B shows the spike patterns of the first and last
layers. To better show the similarity, the spike time of the last
layers was temporally shifted so that the spiking activity in these
two layers could be viewed closer together. The figure demonstrates
that the precise spike pattern was well preserved even until the final
layers.  To quantify the similarity of two spike trains (sums of Dirac
delta functions) $S_1(t)$ and $S_2(t)$, we define the inner product
$\<S_1,S_2\>=\int_0^T\int_0^T S_1(t)D(t-t')s_2(t') dt dt'$, using the
entire duration of simulations, $T$, and a smoothing kernel
$D(t)=\exp(-t^2/(2\tau_D^2))/\sqrt{2\pi\tau_D^2}$ with $\tau_D=10$
ms. The overlap of the two spike trains are then measured, using the
correlation coefficient, by
$\<S_1,S_2\>/\sqrt{\<S_1,S_1\>\<S_2,S_2\>}$. The overlap of the
external input pulses and the spiking activity in different layers are
compared more systematically in Fig. \ref{fig:LIF}C for various values
of the feedforward synaptic strengths, $w$. Note that the amplitude of
external input pulses to the first layers of the chains was always
fixed at 10 mV.  The key parameter for the signal propagation was the
effective input amplitude, $n w$, which is the product of synaptic
strengths and the number of neurons in each layer. When this effective
coupling strength was too small, the activity could not be
successfully propagated to the next layers, and when the effective
coupling strength was too strong, even a spontaneous firing of a
single neuron was sufficient to activate most of the neurons in the
next layer.  For the network structure explored here, the best overlap
was achieved using about 1 mV of feedforward synaptic strength. The
figure suggests that the fine-tuning of the synaptic strength is not
critical for the memory lifetime because the difference between the
overlaps in the 10th and 20th layers did not expand rapidly as
mistuning from the optimal parameter value increased.

\section{Discussion}

I estimated the memory lifetime achieved by coupled nonlinear
neurons. In contrast to the previously proposed semi-linear scheme
that provided the order $\sqrt{N}$ memory lifetime
\cite{Ganguli2008}, I have shown that an order $N/\log N$ memory
lifetime can be achieved by appropriately using nonlinear dynamics.
The derived asymptotic scaling was invariant to the accuracy of the
information buffered. The proposed nonlinear network outperformed a
previously proposed semi-linear scheme in a wide range of parameters,
in particular, with a large number of neurons and large noise.  I have
also demonstrated that the previously proposed semi-linear scheme is
sensitive to the noise level, \ie, a small increase in the noise level
causes monotonic decay of the average activity, turning the order
$\sqrt{N}$ memory lifetime to order $1$. The nonlinear scheme proposed
in this paper, on the other hand, uses large gain to prevent the
activity from decaying and to alleviate the accumulation of noise
using error-correcting nonlinear dynamics.  Because the mathematical
model studied here is general, the result that a network is capable of
buffering sequential input much longer than individual elements is
potentially applicable to
other systems beyond neural networks, such as, gene/protein and social networks.

We considered in this paper the sequential memory task that aims to
reconstruct a whole dynamical sequence of input after some delay. Note
that this task is different from delayed matching working memory tasks
\cite{Fuster1973,Goldman-Rakic1995}, where a brief stimulus is
presented only at a certain time. The major difference is that stimuli
presented at different timings can interfere with each other under the
sequential memory task. This typically happens when recurrently
connected networks are used to buffer a sequence of input
\cite{Busing2010,Lim2011}. For example, under the sequence
  generations by Hopfield-type networks
  \cite{Hopfield1982,Sompolinsky1986,Kleinfeld1986} or by winnerless
  competition networks \cite{Seliger2003,Bick2009}, the activity
  converges to one of the learned patterns, and the presentation of a
  new pattern disrupts the current state. Hence, the delay-line
  structure is often considered for a sequential memory task to
  prevent the interference of signals presented at different timings
  \cite{Ganguli2008}.  While nonlinear attracting dynamics has been
  utilized for non-sequential working memory tasks
  \cite{Camperi1998,Lisman1998,Koulakov2001a,Goldman2003}, this study
  shows that its error-correcting property also provides long-lasting
  memory for a sequential memory task with a feed-forward network architecture.

The feedforward network structure presented in this paper was studied
in the context of synfire chains
\cite{Abeles1991,Aertsen1996,Diesmann1999,Rossum2002,Vogels2005,Kumar2008},
where precise temporal patterns of spikes are their prominent
characteristic. Temporally precise spiking patterns have been
 observed across different brain areas and different
recording conditions
\cite{Takahashi2010,Hahnloser2002,Pastalkova2008,Ji2007,Ikegaya2004,Jin2007}.
There is also some experimental evidence suggesting that the
synfire chain is the underlying network architecture in the brain for
generating precise temporal sequences \cite{Long2010}.  Although the
effect of noise on the gain was systematically studied
\cite{Herrmann1995}, the contribution of occasional large noise that
blocks or spontaneously ignites synfire activity
\cite{Bienenstock1995,Tetzlaff2002} was not theoretically
analyzed. In particular, the trade-off between the length of the chain
and the reliability of activity being propagated for a fixed total
number of neurons was not elucidated. I demonstrated that such
occasional large noise prevents synfire chains from achieving an
extensive memory lifetime, and the resulting $\sim
N/\log N$ memory lifetime is the direct consequence of such noise.

Reservoir computing \cite{Jaeger2009,Maass2002} was recently proposed
as an attractive paradigm for universal and dynamical
computation. This is one candidate network that can also perform
sequential memory tasks \cite{White2004}. According to this paradigm,
dynamical input is provided to a pool of neurons, called a reservoir,
which buffers the history of the input and extracts many useful
features of the input sequence.  Some linear readout units are placed
on top of this reservoir and trained for a specific task, for example
for reconstructing past input, while the reservoir itself remains
task-nonspecific. 
One of the fundamental aspects of 
reservoir computing is that a reservoir buffers past input sequences
so that the readout unit can successfully combine the history of the
input stimuli.
Although randomly connected networks are
  commonly used as the reservoir, the optimal structure of the
  reservoir is not yet known \cite{Lazar2009}. The current study has 
  shown that a feedforward structure is suitable to buffer sequences
  of events. Despite its benefit for sequential memory, making the
  whole network into a feedforward network is probably not a good
  idea.  In addition to memory, it is also important for the reservoir
  to map input to a high dimensional ``feature space'' so that a
  linear readout has access to useful features
  \cite{Jaeger2009,Maass2002,Sussillo2009,Bertschinger2004,Busing2010}.
The current study suggests instead that it would be a promising approach to
embed feedforward chains with high gain as a memory-specific
sub-network for a wide class of tasks that requires a long-lasting
sequential memory. This guarantees a memory lifetime that is nearly
proportional to the size of that memory-specific sub-network.  We
should note that a straightforward implementation of randomly and
recurrently connected nonlinear neurons only achieved a memory
lifetime of order $\log N$ \cite{Busing2010}. In view of the fact that
the best possible scaling of memory lifetime is $\sim N$ for
non-sparse input sequences \cite{Ganguli2010}, it is clear that the
error-correcting feedforward network studied in this paper with $\sim
N/\log N$ memory lifetime is a promising candidate for general
dynamical computations requiring a recent history of activity.

\section*{Acknowledgements}
The author would like to thank L. F. Abbott, Surya Ganguli, Xaq
Pitkow, Fred Wolf, and Alex Ramirez for their helpful comments. T. T. was
supported by the Special Postdoctoral Researchers Program of RIKEN.

\end{document}